\documentstyle[sprocl,epsfig]{article}

\def\bex{\begin{equation}}
\def\eex{\end{equation}}
\def\bea{\begin{eqnarray}}
\def\eea{\end{eqnarray}}
\def\lab#1      {\hbox{\small #1} }
\newcommand{\be}{\begin{eqnarray}}
\newcommand{\ee}{\end{eqnarray}}
\newcommand{\ben}{\begin{eqnarray*}}
\newcommand{\een}{\end{eqnarray*}}
\newcommand{\la}{\langle}
\newcommand{\ra}{\rangle}
\newcommand{\half}{\frac{1}{2}}

\newcommand{\stbi}{\alpha_{\hat{\partial} b_{i}}}

\def\mb#1         {\mbox{\boldmath $#1$}}
\def\diffn#1	  {\Delta^{-}_{#1}}

\begin{document}

\title{SIMULATIONS IN $SO(3) \times Z(2)$ LATTICE GAUGE THEORY}

\author{Andrei Alexandru and Richard W. Haymaker\footnote{Talk presented
at Confinement 2000, Osaka, March 7 - 10 by R. Haymaker}}

\address{Department of Physics and Astronomy, 
 Louisiana State University, \\
Baton Rouge, Louisiana 70803-4001, USA\\
E-mail: alexan@rouge.phys.lsu.edu, haymaker@rouge.phys.lsu.edu}

\maketitle\abstracts{We explore simulations on periodic lattices 
in the Tomboulis $SO(3) \times Z(2)$ formulation.
We measure gauge invariant vortex counters for ``thin", ``thick" and ``hybrid" vortex sheets in order to tag Wilson loops by the occurance
of gauge invariant vortices linking them.  We also measure projection vortex counters defined in the maximal center gauge for comparison.
}

\section{Introduction}
\label{Introduction}

Lattice QCD continues to maintain an important role in the search
for the physics of color confinement.  The lattice regulator maintains gauge invariance
at all costs.  Dynamical variables are group elements rather than elements of a Lie algebra.
As a consequence many of the topological features that are prominent candidates 
for elucidating the physics of confinement have natural lattice definitions.  These include
U(1), Z(N), SU(N)/Z(N) monopole loops, Dirac sheets, Z(N) and SU(N) vortex sheets etc.  
These objects are often abundant in  U(1) and SU(N) lattice gauge theories.
They become singular only as one approaches the continuum limit.

Yaffe \cite{y}, Tomboulis \cite{t} and Kovacs and Tomboulis \cite{kt}
have developed a formulation of SU(N) gauge theory that is manifestly   SU(N)/Z(N) 
invariant.  In this formulation, center elements, Z(N), multiplying each link leave the
action and measure invariant. New Z(N) variables, defined on 
plaquettes, $\sigma(p)$,  carry the Z(N) degrees of freedom.
This  formalism, equivalent to the standard SU(N) form,  
allows an elegant topological 
classification of the SU(N)/Z(N) and Z(N) vortex 
configurations occuring on the lattice.  See also Refs. 4 - 7.

This talk is a report on our  paper \cite{ah}, in which 
we address the issues of simulations in the Tomboulis variables on a periodic
lattice.  Many results have followed from this formulation without doing simulations \cite{ttt}.  
As a first calculation, we tag Wilson loop measurements 
by the occurance of vortices linking the loop using a number of different vortex counters.  We restrict our attention to SU(2).

We also measure the P (projection) vortex counter in the original SU(2) formulation %following \cite{dfgo,fgo,agg,lr,lrt,elrt,afe,montero,stephenson} 
following Refs. 10 - 18 for comparison.   Projection vortices arise in a Z(2) gauge 
theory derived from the original SU(2) theory by going to the maximal center gauge  and then replacing links  by 
$\lab{sgn} ( \lab{tr} ( U_{x, \mu}))$.
The projected theory has ``thin" Z(2) vortices defined on one lattice
spacing.  They have been found to be 
correlated with center vortices and therefore a 
measurement of P vortices is a predictor of them.

\section{$Z_{Z(2) \times SU(2)/Z(2)}$ on a torus }
\label{Z}

The Wilson form of the partition function,
\be
Z 
=
\int 
\left[
  d U(b) 
\right]
\exp
\left(
   \beta 
   \sum_{p}
   \half
   \lab{tr} [U(\partial p)]
\right),
\label{wilsonaction}
\ee 
can be cast into a form in which the variables transform under 
$Z(2)$ and $SU(2)/Z(2) = SO(3)$ separately \cite{t,kt,ah}.
\be
Z  
&=&
\int 
\left[
     d U(b)
\right]
\sum_{\sigma(p)}
%\nonumber 
%\\ 
%&& 
C[\sigma(p)\eta(p)] 
\exp
\left(
   \beta 
   \sum_{p} 
   \half|\lab{tr} [  U(\partial p)]| 
   \sigma(p)
\right),
\label{z36}
\ee
where
\be
C[\sigma(p)\eta(p)] 
= 
\sum_{\tau(p)}
   \prod_{p}  
   \chi_{\tau(p)}\left(\sigma(p)\eta(p) \right)
   \prod_{b}        
   \delta \left( \tau(\widehat{\partial} b \right).
\label{CC}
\ee

The derivation makes use of the invariance of the Haar measure under
$Z(2)$ transformations $U(b) \rightarrow U(b) \gamma(b)$ of the links. Summing over $\gamma(b)$, the link matrices become $SU(2)$ representatives
of the invariance group $SU(2)/Z(2) = SO(3)$. The $Z(2)$ dependence is transferred to new  variables defined on plaquettes
$\sigma(p)$.  The variable $\eta(p)$ is the sign of the plaquette
$\lab{tr} [U(\partial p)] = |\lab{tr} [U(\partial p)]| \times \eta(p)$.

$C[\sigma(p)\eta(p)]$ is a constraint, equal to either a positive constant or zero. 
The $Z(2)$ character, $\chi_{\alpha}(\beta) = \pm 1$, where $\alpha = \pm 1$ 
labels the representation and $\beta = \pm 1$ is the argument. 
$\chi_+(+) = \chi_+(-) = \chi_-(+) = - \chi_-(-) = +1$.
The delta function constrains the $\tau$ summation to contribute only if 
 an even number
of negative $\tau$ plaquettes are in the coboundary of each link.  

For free boundary conditions  \cite{t}
\ben
C[\alpha(p)]=
\prod_{c}
\delta\left(\alpha(\partial c)  \right)
\times \lab{constant}
\een
This requires that there be an even number of 
$\alpha(p) = \sigma(p) \times \eta(p)$
plaquettes on the faces of all elementary cubes.

\section{Zero weight configurations on the torus}
\label{zero_weight}

This constraint $C[\alpha(p)]$ is further restrictive for periodic boundary
conditions. It gives zero for a
vortex configuration of negative plaquettes on a coclosed surface 
wrapping around the torus.  This means that one can not excite
a configuration corresponding to the topologically stable anti-periodic
boundary conditions and vice versa. 

To see this we use a result from Ref. \cite{ah} Note that the $\{\tau\}$
configurations form a group with multiplication rule
$(\tau_1 \tau_2)(p) = \tau_1(p) \tau_2(p)$.  Using the invariance of the 
group sum over $\tau$ we can write
\ben
C[\alpha(p)] 
&=& 
\sum_{\tau(p)}
   \prod_{p}  
   \chi_{\tau(p)}\left(\alpha(p) \right)
   \prod_{b}        
   \delta \left( \tau(\widehat{\partial} b )\right), 
\label{CCC}
\\
&=& 
\sum_{\tau(p)}
   \prod_{p}  
   \chi_{\tau_0(p) \tau(p)}\left(\alpha(p) \right)
   \prod_{b}        
   \delta \left( \tau(\widehat{\partial} b )\right),
\\
&=& \la \tau_0, \alpha \ra
 \times C[\alpha(p)],
\label{zeroo}
\een
where $\la \tau, \alpha \ra \equiv
\left\{
   \prod_{p}   
   \chi_{\tau(p)}\left(\alpha(p)\right)
\right\}$.  
If we can find a group element
$\tau_0$ for which $ \la \tau_0, \alpha \ra \ne 1$ then $C[\alpha]  = 0$. 

Take the example of a coclosed stack of $\alpha = -1$ (1,2) plaquettes
at locations $(0, 0,  \xi, \eta)$ and
a closed tiled surface of $\tau = -1$ (1,2) plaquettes at locations 
$(\mu, \nu, 0, 0)$.  They have one plaquette in common, giving 
one factor of $-1$ in $\la \tau_0, \alpha \ra
$, and hence $C[\alpha] = 0$.

As a second example let us consider how $C[\alpha]$ gives the cube constraint.
Let us suppose that a particular cube violates this constraint with an odd number of faces
 with $\alpha = -1$.  Then take  a configuration $\tau_0$ 
which takes values $-1$ on all 6 faces of this particular cube.  This is a closed
surface and therefore satisfies the constraints imposed on $\tau$.  Again
$   \prod_{p}      \chi_{\tau_0(p)}\left(\alpha(p)\right) = -1  $
and therefore  $C[\alpha] = 0$.

\section{$\sigma(p)$ configuration space}
\label{simulation}
We are interested in 
simulating in the variables $\{U(b), \sigma(p) \}$.  
Allowed configurations $\alpha(p)$ are defined indirectly by the constraint, function $C[\alpha]$.  
The corresponding simulation would
be cumbersome to implement.

We proposed \cite{ah} a constructive definition of allowable $\alpha(p)$ configurations by building
them up from ``star transformations", i.e. correlated sign flips of the $\sigma $ plaquettes 
occurring in the co-boundary of each link. 
We showed \cite{ah} that the constrained updates of six plaquettes
reaches all allowed  $\{ \sigma \}$ configurations, and in fact is 
identical to the above definition.
In this section we summarize this result.

Before discussing the $\sigma$ configurations let us first describe the link updates.
This is a straightforward generalization of the link updates for $SU(2)$.   
The proposed
change in a link might change the sign of the $\eta$ plaquettes in the co-boundary of 
the link.  If one of these changes sign, we need to flip the sign of the 
corresponding $\sigma$ plaquette so that the $\alpha$ configuration is unchanged.
Then the Monte Carlo step is essentially the same as for the SU(2) update. 

Next consider the above mentioned star transformations.   Our proposed 
update is to flip the sign of the six $\sigma$ plaquettes forming the co-boundary of 
the links.

It is easy 
to see that both these update steps will preserve the local constraints, i.e. the cube constraints. 

Next examine the global constraints. Consider
the operator constructed out of $\mu, \nu$ plaquettes \cite{kt}
\ben
{\cal N}_{\mu, \nu}  =  \prod_{p\in S_{\mu, \nu}} 
\eta(p) \sigma(p)  \equiv  
 \eta(S_{\mu, \nu})  \sigma(S_{\mu, \nu}) = \pm 1,
\een
where $S_{\mu, \nu}$ is a whole tiled $\mu, \nu$ plane.  
${\cal N}_{\mu, \nu} = \pm$ 
for an even/odd number of vortices of stacked $ \mu, \nu $ plaquettes wrapping 
around the orthogonal $\xi, \eta$ directions of the torus. 

We start with ${\cal N}_{\mu, \nu} = +1$  in all 6 planes. 
It is easy to see that our update algorithm preserves 
${\cal N}_{\mu, \nu}$.

Let us define the relevant sets of configurations more carefully.
\be
C[\alpha]
&\equiv& 
\sum_{\tau(p)}
   \prod_{p}  
   \chi_{\tau(p)}\left(\alpha(p)\right)
   \prod_{b}        
   \delta \left( \tau(\widehat{\partial} b) \right), 
\nonumber
\\
&=& 
\sum_{\tau \in {\cal C}}
   \prod_{p}   
   \chi_{\tau(p)}\left(\alpha(p)\right),
\nonumber
\\
&\equiv& 
\sum_{\tau \in {\cal C}}
\la  \tau, \alpha \ra \;\;\; = \;\;\;
\sum_{\tau \in {\cal C}}
\la  \alpha, \tau \ra.
\label{CCCC}
\ee

The bracket, $\la  \alpha, \tau \ra = \pm 1$, is negative if and only if 
there are an odd number of plaquettes for which 
both $\alpha(p) = -1$ and $\tau(p) = -1$.

The second line is an alternative way to specify the sum, where:
\ben
{\cal C}
= 
\{\tau\in{\cal A}| \prod_{b} \delta \left( \tau(\widehat{\partial} b) \right)=1\}.
\een

${\cal C}$ is a subgroup of the group  ${\cal A}$
of all configurations,  $|{\cal A}| = 2^{6 N}$ in number, where
$N$ is the number of lattice sites. 
{\em ${\cal C}$ is the group of all configurations $\{\tau\}$ with an
 even number of $\tau = -1$ 
plaquettes 
occurring in the co-boundary of every link, i.e. 
forming a closed tiled  surface  of negative plaquettes.} 

There is a second group of interest, 
\ben
\overline{\cal C}=\{\alpha \in{\cal A}|\la\alpha,\tau \ra 
=1 \;\;\;\;\forall\tau\in{\cal C}\}.
\een
{\em
${\overline{\cal C}}$ 
is the group of all configurations 
$\{ \alpha \}$ 
for which
$C(\alpha)$ is different from zero.  
}

Therefore ${\overline{\cal C}}$ is the group of $\alpha$ 
configurations which have non zero weight 
in the partition function, Eqn.(\ref{z36}).
Further,  ${\cal C}$ is the group of $\tau$ configurations that form 
closed tiled surfaces as required by the explicit constraints in  Eqn.(\ref{CCCC}).

The group ${\cal C}$ has only an implicit definition here.  The group ${\overline{\cal C}}$
has an implicit definition in terms of this ${\cal C}$. Therefore its definition 
is even more indirect.  Even without an explicit definition,
we have been able to specify precisely those configurations $\{ \alpha \}$ that contribute
non-zero weight to the partition function.   

There is a third group of interest,
\ben
{\cal D}=\{\alpha\in{\cal A}|\alpha=\prod \stbi \}.
\een 
where $\stbi$ refers to an individual star transformation on the 
$i$'th link $b_i$, and the product indicates
all possible products of them.
{\em
${{\cal D}}$ 
is the group of all configurations 
$\{ \alpha \}$ 
which can be built out of products 
of ``star transformations" starting from the
identity configuration.
}

This is the constructive definition that is straightforward to implement in a simulation.
We show in Ref. \cite{ah} that 
the group $\cal D$ is identical to the group ${\overline{\cal C}}$.  In this way we have shown that by our proposed algorithm
is ergodic.

\section{Simulation of  vortex counters}

Vortices have long been considered as prime candidates for the
essential dynamical variable to describe confinement.    A simulation offers
a tool that allows one to correlate the occurance of vortices with values of other dynamical variables. Hence as 
a first application we use this formalism to measure various vortex counters for
Wilson loops.

In the $SO(3) \times Z(2)$ formulation the Wilson loop is given by 
\ben
W[C] =  \la \half \lab{tr} [C] \eta_S \sigma_S \ra_{C = \partial S},
\label{wilson}
\een
where $\eta_S$ and $\sigma_S$ are products of $\eta$ and $\sigma$ over any 
spanning surface \cite{y,t,kt}.

Kovacs and Tomboulis \cite{kt} define three  vortex counters for 
thick vortex sheets, thin vortex sheets, and hybrid (patches 
of each on the sheet).

\begin{figure}[h]
\epsfig{file=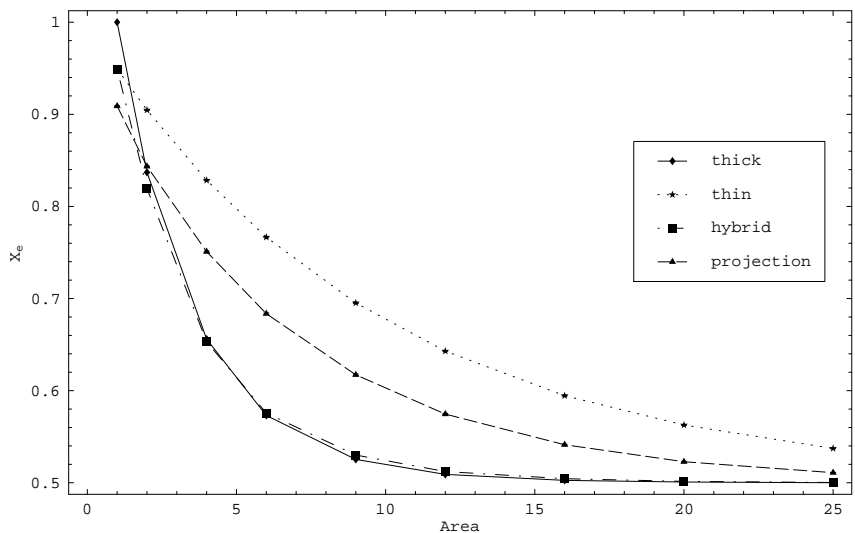,width=11.9cm}
\caption{Fraction of Wilson loops with an even number of 
vortices piercing the minimal surface, $x_e$. ($x_o=1-x_e$)}
\end{figure}

\begin{figure}[h]
\epsfig{file=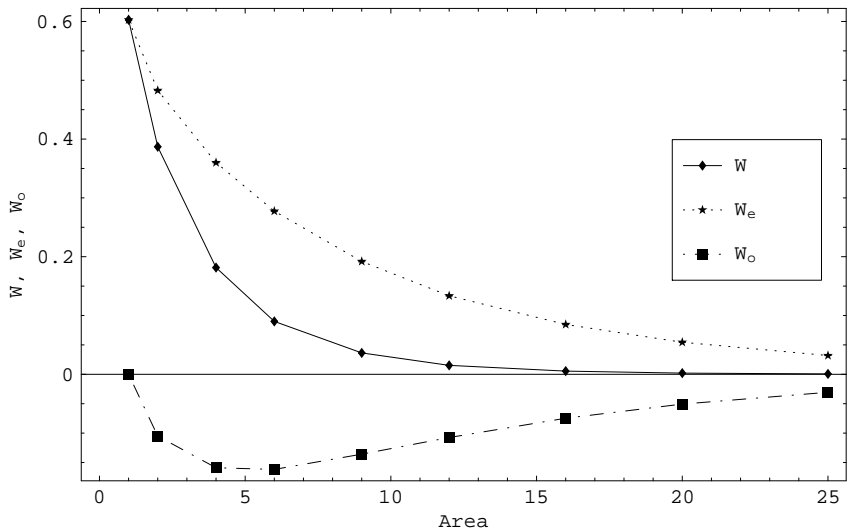,width=11.9cm}
\caption{Wilson loop, W, and tagged Wilson loop with even or 
odd number of  ``thick" vortices piercing the minimal surface.}
\end{figure}

{\bf Thin:}
\ben
{\cal N}_{\lab{thin} } \equiv  \sigma_{S}.
\een
If this value, $ = \pm 1$, is independent of the spanning surface, then 
this counts thin vortices.

{\bf  Thick:}
\ben
{\cal N}_{\lab{thick} } \equiv \eta_{S} \; \lab{sgn}[tr W(C)].
\een
This object is counting something more elusive since unlike the above case, the  vortex structure is spread over many lattice spacings.  
Nevertheless 
it is always possible to find a representative of SO(3) such
that the $\eta$  vortex defines the 
topological linkage \cite{kt}. 
The  $\eta$  
vortices can be deformed by a $Z(2)$ transformation of links 
giving  different representatives of SO(3) without cost
of action.  
One can move a linked $\eta$ vortex sheet so that it no 
longer links the Wilson loop and further even transform it away.  
However in this case the negative contribution will be transferred
to one of the perimeter links  of the Wilson loop, and it will not affect the value of  ${\cal N}_{\lab{thick} }$.  Again if this is independent
of the spanning surface, then this counts  ``thick" vortices.

{\bf Hybrid:}
\ben
{\cal N}_{\lab{hybrid} } = {\cal N}_{\lab{thin} } \times
{\cal N}_{\lab{thick} } =   \sigma_{S} \eta_{S} \;\lab{sgn}[tr W(C)].
\een
As one considers all spanning surfaces, the sign of 
${\cal N}_{\lab{thin} }$ might change. However
if the sign of ${\cal N}_{\lab{thick} }$ always compensates then 
this counts hybrid vortices.

\section{Numerical Results}

\begin{figure}[h]
\epsfig{file=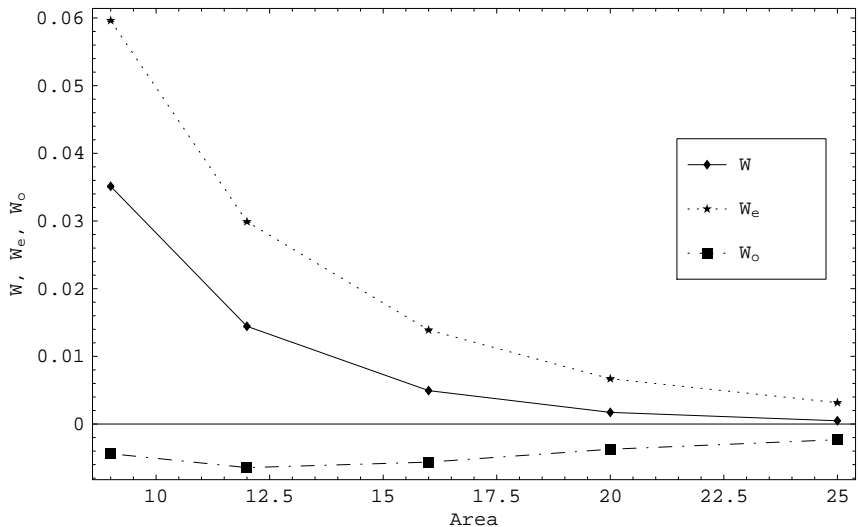,width=11.9cm}
\caption{
Wilson loop, W, and tagged Wilson loop with even or 
odd number of  projection vortices piercing the minimal surface.
}
\end{figure}

\begin{figure}[h]
\epsfig{file=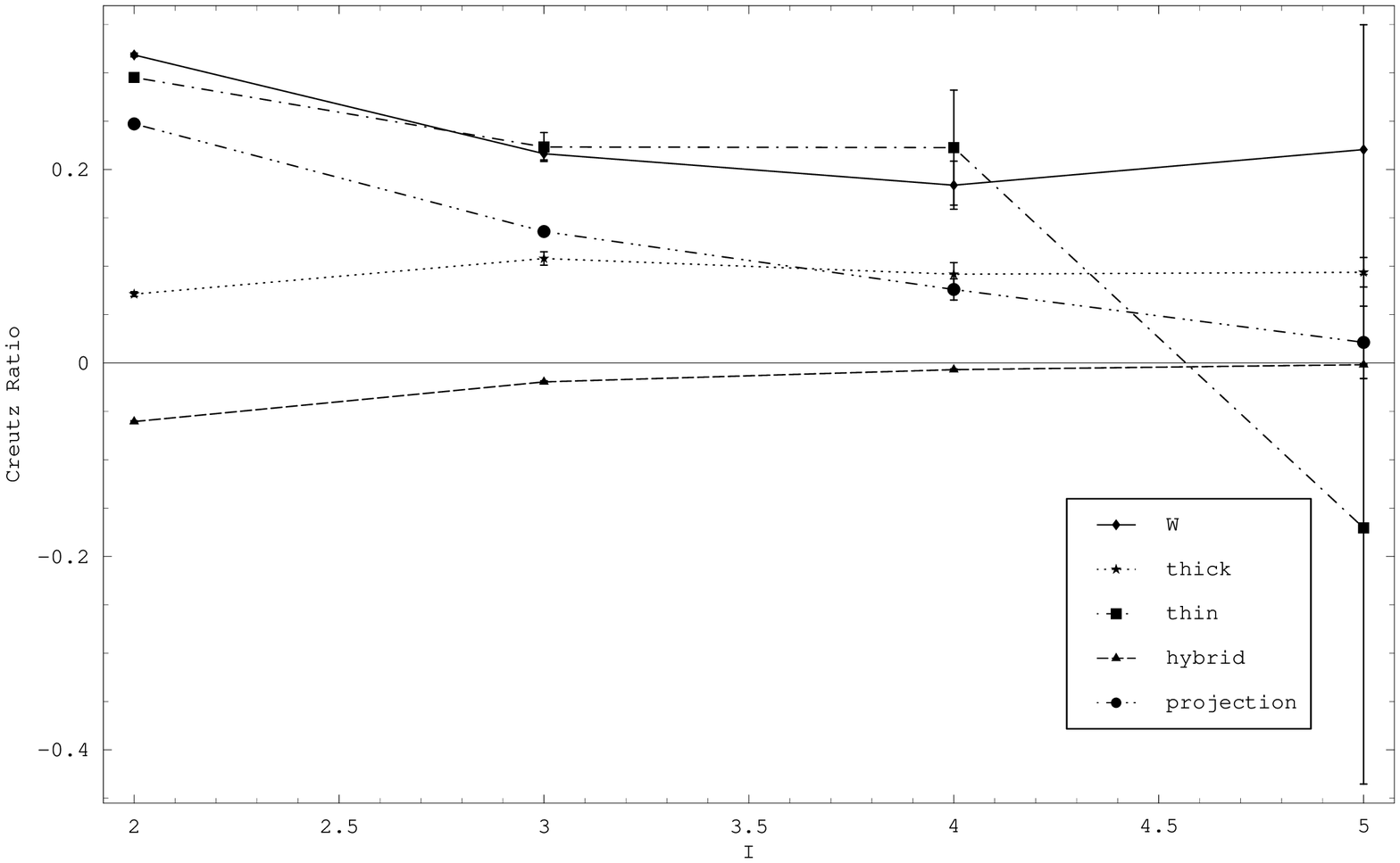,width=11.9cm}
\caption{ Creutz Ratio: $\chi (I,I)=- \ln \frac{W(I,I) W(I-1,I-1)}{W(I,I-1)^2}$, similarly 
for tagged Wilson loops $W_e$.}
\end{figure}

Simulations were done on a $12^4$ lattice for $\beta = 2.30$. 
 Measurements were binned
to 10 bins and jackknife errors calculated.
[thin]: 200 ;
[thick]: 400 ;
[hybrid]: 200 ;
[projection]: 1000 measurements.
We monitor the coincidence of Z(2) and SO(3) monopoles which can
slip due to round off error.

Kovacs and Tomboulis' definitions require measurements on {\em all} spanning surfaces.  
We measure here only the minimum spanning surface. 
Then, for example, 
${\cal N}_{\lab{thin} } = -1$ does not distinguish
thin from hybrid, and similarly ${\cal N}_{\lab{thick} } = -1$ does not distinguish thick from 
hybrid. 

{\bf Fig. 1} shows the fraction of Wilson loops 
which have an even number of vortices, $x_e$,
as a function of the area.
All the counters 
approach $50 \%$ from above with similar behavior ($x_o = 1 - x_e$).
They are each counting different things and 
are not expected to be equal.   

${\cal N}_{\lab{hybrid} }$ is a good reference curve since it 
measures the sign of the Wilson loop itself.  The area law arises from a  near cancelation of fluctuating values due to approximately 
equal occurance and absence of thin, thick or hybrid vortices.

The ``thick" curve lies on top of the ``hybrid" one. 
Hence the added factor of  $\sigma_S$ in the hybrid counter has little effect here.  We expect $\sigma$ vortices to be heavily suppressed
for increasing $\beta$ since they cost action proportional to the vortex area.   However at $\beta = 2.30$ the density of Z(2) (or SO(3)) monopoles 
$ = 0.2155(2)$  (Random plaquette signs would give a density of $ ~ 0.5$). 
Hence in spite of the near coincidence of these two curves, $\sigma$ vortices  and $\sigma$ patches of hybrid vortices  are important at this value of  $\beta$.  There is further evidence below.
The ${\cal N}_{\lab{projection} } $ 
case gives the same result as reported in Ref \cite{dfgo}.

{\bf Fig.2
[Thick segment piercing the minimal surface] }
By definition $W_e$ has an even number of thick segments piercing the minimal
surface.  Yet it still has an exponential fall off with area.  
See Fig. 4 which gives the Creutz ratio showing that $W_e$
has about half the string tension of the $W$.  The thin segments
are still present and they account for the disordering.

{\bf Fig. 3  [Projection vortices piercing the minimal surface]}
 For loops of area 9 and higher, the 
sign of the vortex counter correlates with $W_o$. These
data agree with Ref. \cite{dfgo}.  
Comparing this with  Fig. 2 [thick] 
the latter data are about a factor of 10 larger, indicating a 
large discrepancy in these two methods of tagging center vortices.

{\bf Fig. 4    [Creutz Ratio]}
$W$ and $W_e$ [thick] show a constant string tension for larger loops.  We suspect that with better statistics, $W_e$ [thin] will also.  
$W_e$ [hybrid] shows the vanishing of the string tension if one removes
the disordering mechanism completely.  Larger loop areas are needed
to decide if the Creutz Ratio of  $W_e$ [projection] will
also go to zero, or stabilize which would indicate that a disordering mechanism remains.

\section{Summary}

The Tomboulis formalism provides a precise way to separate the physical effects of thin and thick vortices.   Simulations in these 
variables allows one to disentangle these effects as one approaches
the continuum limit where thick vortices are expected to dominate.

\section*{Acknowledgments}
We are pleased to thank E. T. Tomboulis, S. Cheluvaraja and 
P. deForcrand for 
helpful discussions.  This work was supported in part by United States Department of Energy grant DE-FG05-91 ER 40617.

\section*{References}


\begin{thebibliography} {99}
%
\bibitem{y} L. G. Yaffe, Phys. Rev. {\bf D 21}, 1574 (1980).
%
\bibitem{t} E. Tomboulis, Phys. Rev. {\bf D 32}, 2371 (1981).
%
\bibitem{kt} T. G. Kovacs and E. Tomboulis  Phys. Rev. {\bf D 57}, 4054 (1998)
%
\bibitem{mp} G. Mack and V. B. Petkova, Annals of Physics {\bf 123}, 442 (1979);
{\bf 125}, 117 (1980); Z. Phys. {\bf C 12}, 177 (1982).
%
\bibitem{hs} G. Halliday and A. Schwimmer, Phys. Lett. {\bf B 101}, 327 (1981); 
{\bf B 102}, 337 (1981).
%
\bibitem{yo} T. Yonewa   Nucl. Phys. {\bf B 203} [FS5], 130 (1982).
%
\bibitem{c}  J. M. Cornwall, Phys. Rev. {\bf D 26}, 1453 (1979)
%
\bibitem{ah} A. Alexandru and R. W. Haymaker, hep-lat/0002031, to be published in Phys. Rev.
%
\bibitem{ttt} T. G. Kovacs and E. Tomboulis hep-lat/0002004, 9912051, 9908031,
Phys. Lett. {\bf B 463} 104 (1999); Nucl. Phys. B, Proc. Suppl {\bf 73}566 (1999);
Phys. Lett. {\bf 443} 239, (1998); J. Math. Phys. {\bf 40}, 4677 (1999).
%
\bibitem{dfgo} L. Del Debbio, M. Faber, J. Greensite and  S. Olejnik, Phys. Rev. 
{\bf D 55}, 2298 (1997), hep-lat/9802003,
%
\bibitem{fgo} M. Faber, J. Greensite and  S. Olejnik, JHEP 9901:008,1999; 
JHEP 9912:012,1999; hep-lat/9911006; hep-lat/9912002
%
\bibitem{agg} J. Ambjorn, J. Giedt, J. Greensite, hep-lat/9907021, hep-lat/9908020
%
\bibitem{lr} K. Langfeld and H. Reinhardt, Phys. Rev. {\bf D 55},7993 (1997)
%
\bibitem{lrt} K. Langfeld, H. Reinhardt and O. Tennert, Phys. Lett. {\bf B 419},
317 (1998);
%
\bibitem{elrt} M. Engelhardt, K. Langfeld, H. Reinhardt and 
O. Tennert, Phys. Lett. {\bf B 431}, 141 (1998); {\bf 452}, 301 (1999); Phys. Rev.
{\bf D 61}:054504,2000; hep-lat/9908026
%
\bibitem{afe} P. de Forcrand and M. D'Elia, hep-lat/9907028; hep-lat/9909005.
%
\bibitem{montero} A. Montero, hep-lat/9907024.
%
\bibitem{stephenson} P.  W. Stephenson, hep-lat/9909022.
%
\end{thebibliography}
\end{document}